\documentclass[twocolumn,showpacs,preprintnumbers,superscriptaddress,amsmath,amssymb]{revtex4-1}

\usepackage{epsfig}
\usepackage{float}
\usepackage{graphicx}
\usepackage{dcolumn}
\usepackage{bm}
\usepackage{times}
\usepackage{xcolor}
\usepackage{url}

\begin{document}



\title{Self-awareness based resource allocation strategy for containment of epidemic spreading}
\author{Xiaolong Chen}
\affiliation{School of Economic Information Engineering, Southwestern University of Finance and Economics, Chengdu 611130, China}
\author{Quanhui Liu}
\affiliation{College of Computer Science, Sichuan University, Chengdu 610065, China}
\author{Ruijie Wang} \email{ruijiewang001@163.com}
\affiliation{A Ba Teachers University, Aba 623002, China}
\author{Qing Li}
\affiliation{School of Economic Information Engineering, Southwestern University of Finance and Economics, Chengdu 611130, China}
\author{Wei Wang}
\affiliation{Cybersecurity Research Institute, Sichuan University, Chengdu 610065, China}
\date{\today}

\begin{abstract}
Resource support between individuals is of particular importance in controlling or
mitigating epidemic spreading, especially during pandemics.
Whereas there remains the question of how we can protect ourselves from being
infected while helping others by donating resources in fighting against the
epidemic. To answer the question, we propose a novel resource allocation model
by considering the awareness of self-protection of individuals. In the
model, a tuning parameter is introduced to quantify the reaction strength of
individuals when they are aware of the disease.
And then, a coupled model of resource allocation and disease spreading is proposed to study the impact of self-awareness on resource allocation and, its impact on the dynamics of epidemic spreading.
Through theoretical analysis and extensive Monte Carlo simulations,
we find that in the stationary state, the system converges to two states: the
whole healthy or the completely infected, which indicates an abrupt increase in the prevalence when there is a shortage of resources. More importantly,
we find that too cautious and too selfless for the people during the outbreak of an epidemic are both not suitable for disease control. Through extensive simulations, we find the optimal point, at which there is a maximum value of the epidemic threshold, and an outbreak can be delayed to the greatest extent.
At last, we study further the effects of network structure on the coupled
dynamics. We find that the degree heterogeneity promotes the outbreak of disease,
and the network structure does not alter the optimal phenomenon in behavior
response.
\end{abstract}

\maketitle

\textbf{}
\section{Introduction} \label{sec:intro}

Controlling the outbreak of epidemic spreading is one of the most important topics in human
history. During the past decades, the onset of several major global health threats such as
the 2003 spread of SARS, the H1N1 influenza pandemic in 2009, and the western Africa Ebola
outbreaks in 2014 have deprived tens of
thousands of lives all around the world \cite{Chan2003Epidemiology,girard20102009,team2014ebola}. In the present, a novel coronavirus (2019-nCoV)
causing severe acute respiratory disease emerged in Wuhan, China. As of February 4, 2020, there have
been 20528 confirmed 2019-nCoV infections reported in 33 provinces and municipalities
\cite{2019-nCoV}. The surge in infections has led to a severe shortage of medical resources. Thousands of
confirmed and suspected cases await treatment. Facing the rapid outbreak of disease,
people all over the country contribute a resource to support
the populations in the epidemic areas, whereas self-protection is also essential.
Thus the immediate problem is how can we protect ourselves from being
infected while helping others in fighting against the epidemic.

A large number of researchers from various disciplines have made efforts to study
the topic of optimal resource allocation in disease suppressing in the past years
\cite{wan2008designing,gourdin2011optimization,lokhov2017optimal,zhao2018virus,li2020functional}.
For example, Preciado et al. ~\cite{preciado2013optimal}
studied the problem of the optimal distribution of
vaccination resources to control epidemic spreading based on complex
networks. They fond the cost-optimal distribution of
vaccination resource when different levels of vaccination are allowed
through a convex framework.
Further they
studied problems of the optimal allocation of
two typical resources in containing epidemic spreading,
namely the preventive resources and corrective resources throughout the
nodes of the network to achieve the highest level of containment
when the budget is given in advance or finding the minimum budget required to control the spreading
the process when the budget is not specified \cite{preciado2014optimal}.
By using geometric programming, they solved these two problems.
Chen et al.~\cite{chen2017optimal} solved the problem of optimal
allocation of a limited medical resource based on mean-field theory.
They found that if the resource quantity that each node could get
is proportional to its degree, the disease can be suppressed to the greatest extent.
The above works considered the problem from a perspective of mathematical. The problem is solved from the premise that both the number of resources and the spreading state of the epidemic are fixed.

However, the real scenario is more complicated than a static
mathematical problem. Multiple dynamical processes always interact
and co-evolve \cite{granell2013dynamical,pastor2015epidemic}, forming a more realistic starting point.
For example, although, the outbreak of 2019-nCoV induce a severe shortage of
food, medical and defense resources such as surgical mask and disinfectant
within a short time, with more and more resources produced by the
healthy people, and the social-support from home and abroad, the situation
is beginning to ease in Wuhan.
More resources could help curb the spread of the disease.
Resources and disease always interact dynamically in the evolution
process. The coevolution of multiple dynamical processes attracts
extensive research in recent years \cite{wang2019coevolution}. B{\"o}ttcher \cite{bottcher2015disease}
studied the coevolution of resource and epidemics, they found a critical
recovery cost that if the cost is above the critical value,
epidemics spiral out of control into ``explosive'' spread.
Chen et al.~\cite{chen2018} studied the effect of
social support from local connections on the spreading dynamics
of the epidemic. They proposed coevolution spreading model
on multiplex networks and found
a hybrid phase transition on networks with heterogeneous
degree distribution. In this multiplex network framework,
Chen et al. ~\cite{chenoptimal} further impact of preferential
resource allocation on social subnetwork on the spreading dynamics
of the epidemic. They found that the model exhibits different
types of phase transitions, depending on the preference value
of resource allocation.

In addition to the physical resource
that can directly mitigate or control the epidemic spreading,
the awareness of the epidemic in populations is another type of
resource. The public can perceive the threat of epidemic
social network through platform and mass media and then take measure
to self-protect. Thus the interplay between awareness and
epidemic spreading is another topic that attracts extensive
research. A mass of works addressed the problem from
different perspectives considering, for example,
the risk perception, behavioral changes
\cite{funk2010endemic,wu2012impact,yang2019suppression}.
Granell et al. ~\cite{granell2013dynamical}
studied the interplay between the spreading of an
the epidemic, and the information awareness on top of multiplex networks.
Leveraging a microscopic Markov chain approach they
found that the threshold of the epidemics has a metacritical
the point from which the onset increases and the epidemics incidence decreases.
Wang et al.~\cite{wang2016suppressing} studied the coevolution mechanisms using both
real online and offline data, and proposed a coupled model on
multiplex network. They found that disease outbreak in contact
the network can trigger the outbreak of information on a communication network, and
found an optimal information transmission rate that
markedly suppresses the disease spreading.

Awareness of epidemic spreading is critically essential in suppressing
disease outbreak, although there is a mass of works about the coevolution
of awareness diffusion and disease spreading, a question remains to
study. Namely, how awareness influences individual behavior during the
outbreak of disease. The more interesting question is, how individuals will decide whether to allocate their resources
after perceiving the local transmission state. To answer these
questions, a novel resource allocation model is proposed in this paper,
which considers the influence of the awareness of each individual.
We consider that healthy individuals in an outbreak of disease
are the source of various resources, they can not only produce
a medical resource such as drugs, surgical mask but also donate
a resource such as funds and food. Whereas when they perceive
the threat of disease in the local area, they would decide whether to
donate resources to infected neighbors. Since the donation behavior
will lead to less resource for self-protection,
and a more significant probability to be infected. Whereas, when they are aware
of the threat of the disease and refuse to donate resources, they will have
more resources for self-protection and a lower probability of being infected.
Moreover, to study the interplay between the resource allocation
and disease spreading, we propose a coupled dynamical model on complex networks.

Further, we adopt the dynamic message-passing method to solve the coupled model theoretically.
First of all, we investigate the influence of awareness on the coupled dynamics of
resource allocation and disease spreading on the scale-free network. To quantify the
reaction strength of healthy individuals to the information of
the local infection state, a parameter $\alpha$ is defined. A larger value of $\alpha$
indicates more sensitivity of an individual to the disease, and lower intention to
denote resources. Through theoretical analysis and numeric simulations, we find
that the system has only two stationary states, namely the absorb state and
the globe outbreak state. With the increase of $\alpha$, the
epidemic threshold first increases and then decreases, which indicates
an optimal value of $\alpha$.
Further, we find the optimal value at
$\alpha_{opt}$ through extensive simulations, at which the disease
can be suppressed to the greatest extent. Then we explain qualitatively
the optimal phenomenon. At last, we investigate the impact of
degree heterogeneity on the coupled dynamics, and find that
the degree heterogeneity does not alter the optimal phenomenon
and the abrupt increase in prevalence with a shortage of resource,
and the epidemic threshold
increases with the decrease of degree heterogeneity, which
suggests that network heterogeneity promotes the outbreak of disease.

\section{Model descritption} \label{sec:model}

\subsection{Epidemic model}
A resource based epidemiological susceptible-infected-susceptible
model (r-SIS) is proposed to describe coupled dynamics of epidemic spreading
and resource allocation on complex network.
Individuals are represented by nodes in the network and an adjacency matrix
$\mathbb{A}$ is introduced to represent the connection between nodes.
If there is an edge between nodes $i$ and $j$,
the element $a_{ij}=1$, otherwise $a_{ij}=0$. According to this scheme,
any individual can be in two different states: susceptible(S) and
infected(I). The infection propagates between each pair of I-state and
S-state neighbors with an infection rate $\widetilde{\lambda}$ in one contact,
which is assumed to depend on whether the S-state nodes to donate resources, see the details in \emph{Sec.~\ref{sec:res}}.
At any time $t$, each I-state node $i$ recovers with a recovery rate $r_i(t)$.
Resource including the medical, funds and food can promote recovery of
patients from disease ~\cite{kulik1989social,nausheen2009social}. Thus we
define the recovery rate of each I-state node as a function of the resource quantity
received from healthy neighbors in this paper.
As each I-state node will get a different amount of resource,
the recovery rate varies from node to node. Consequently, the
the recovery rate of any node $i$ at time $t$ can be defined as
\begin{equation}\label{reci}
 r_i(t)=1-(1-\mu)^{\varepsilon{\omega_i(t)}},
\end{equation}
where $\omega_i(t)$ is resource quantity of node $i$ received from healthy neighbors
at time $t$, and $\mu$ is the basic recovery rate.
A parameter $\varepsilon\in[0,1]$ is introduced in our model to
represent the resource utilization rate ~\cite{mackie2007health}. Since in real scenario, there is
the common phenomenon of the waste on
resource~\cite{jaarsma1999effects,gul2012computer} in medical and
other service systems, implying the resource received from
healthy neighbors may not be fully utilized on curing and
recovery. Without loss of generality, we set $\mu_{r}=0.6$
throughout this work, i.e., only $60\%$ of the resources received
are used.

For the r-SIS model, we define $\rho_i(t)$ is the probability
that any node $i$ is an infected state. The fraction of infected
nodes in a network of size $N$ at time $t$ can be calculated by
averaging overall $N$ nodes:
\begin{equation}
  \rho(t)=\frac{1}{N}\sum_{i=1}^{N}\rho_i(t).
  \label{density}
\end{equation}
Further we define the prevalence of the disease in the stationary state as
$\rho\equiv\rho(\infty)$

\subsection{Resource allocation model based on behaviour response}\label{sec:res}
In the real scenario, healthy individuals can produce resources.
For simplicity, we consider that each individual (node) in the network can
generate one unit resource at a time step.
During an outbreak of a disease, the susceptible individuals
can perceive the threat of the disease intuitively by
acquiring the information from direct neighbors. Generally,
the more infected neighbors of an individual,
the deeper it will be aware of the disease \cite{funk2010modelling,wang2015dynamics}.
People aware of the disease will have different reactions ~\cite{funk2010modelling}.
And to quantify the reaction strength of an individual to the
local information of disease, a tuning parameter $\alpha$ is introduced.
Based on the description above, we can define
the probability that a healthy individual with $m$
infected neighbors donate resource as
\begin{equation}\label{disProb}
  q(m)=q_0(1-\alpha)^{m},
\end{equation}
where $q_0$ is a basic donation probability. When $\alpha=0$,
all healthy nodes have the same donation probability $q_0$.
Besides, we consider that a healthy node will
donate one unit resource equally to its I-state neighbors
at a time.
Based on the resource allocation scheme, the amount of
resource $\omega_{j\rightarrow{i}}$ that node $j$ with
$m$ infected neighbors donate
to one of its I-state neighbor, $i$ can be expressed as
\begin{equation}
  \omega_{j\rightarrow i}=q(m)\frac{1}{m}.
  \label{ResTrans}
\end{equation}

When disease breaks out in the human population, people aware of a disease
in their proximity will take measures to reduce their susceptibility, leading
to a reduction in the effective rate of infection ~\cite{funk2009spread,granell2013dynamical}.
We consider that if an individual is aware and refuse to donate resource
for self-protection, it reduces its infectivity by a factor $c$.
The basic infection rate is denoted as $\lambda$, and the actual rate of
infection is denoted as $\widetilde{\lambda}$, which can thus be expressed
as
\begin{equation}\label{lambda}
\widetilde{\lambda}=
\left\{
     \begin{array}{ll}
       \lambda, & \hbox{if distribute the resources;} \\
       c\lambda,& \hbox{else. }
     \end{array}
   \right.
\end{equation}
If a healthy individual donate resource to infected neighbors,
it has a larger probability to be infected, on the contrary, it
has a relatively smaller probability to be infected. The effective infection rate of
any node $i$ can be also expressed as a function of $q$,
\begin{equation}\label{lami}
  \widetilde{\lambda}_i=q(m)\lambda+[1-q(m)]c\lambda
\end{equation}

\section{Dynamic message-passing method}
In order to theoretically analyze the dynamic processes,
we develop a generated dynamic message-passing method (GDMP)
\cite{karrer2010message,shrestha2015message}.
In this method, the message $\theta_{j\rightarrow i}$
are defined on the directed edges of a network
to carry causal information of the flow of contagion, which
can only transfer one way along directed links.
$\theta_{j\rightarrow i}$ represents the probability that node $j$ is
infectious because it was infected by one of its neighbors
other than node $i$. In computing $\theta_{j\rightarrow i}$, we
only take into account the contributions to $\rho_j$ that come
from the neighbors other than $i$. The higher order process of
$j$ being infected by $i$ and then passes the infection back to
$i$ is neglected. Combine $\theta_{j\rightarrow i}$ and
Eq.~(\ref{disProb}) for resource allocations, the resources
$\omega_{i}(t)$ that an infected node $i$ receives from its healthy
neighbors can be expressed as
%
\begin{equation}
\omega_i(t)=\sum_{j}a_{ij}[1-\theta_{j\rightarrow i}(t)]\frac{q[m_j(t)]}{m_j(t)},
\label{resource}
\end{equation}
where $m_j(t)$ is the expected number of I-state neighbors of node $j$ at time $t$, which is
expressed as:
\begin{equation}\label{mt}
  m_j(t)=\sum_{h\neq{i}}a_{jh}\theta_{h\rightarrow{j}}(t)+1,
\end{equation}
where the plus one takes into account that node $i$ is in infected
at this moment. The factor $(1-\theta_{j\rightarrow i}(t))$ in
Eq.~(\ref{resource}) stands for the probability
that node $j$ is susceptible at time $t$. With the
definition above, the discrete-time version of the
evolution of $\rho_i(t)$ in a time interval $\Delta t$ reads \cite{gomez2010discrete}
\begin{equation}
\rho_i(t+\Delta t)=(1-\rho_i(t))(1-\Omega_i(t))+[1-r_i(t)\Delta
t]\rho_i(t), \label{dynamic}
\end{equation}
where $\Omega_i(t)$ is the probability that node $i$ is not infected by any neighbor with
the product being over the set $\mathcal{N}_{i}$ of the neighbors
of node $i$. The expression of $\Omega_i(t)$ is as follow:
\begin{equation}
\Omega_i(t)= \prod_{j\in \mathcal{N}_{i}}[1-\Delta
t\,\widetilde{\lambda}_i(t)\,\theta_{j\rightarrow i}(t)], \label{q}
\end{equation}
Note that the first term on the right-hand side of Eq.~(\ref{dynamic})
stands for the probability that node $i$ is in S-state
and infected by at least one of its neighbors. The second
term is the probability that node $i$ is in I-state and
does not recover. Similarly, we can get the time evolution of
$\theta_{j\rightarrow i}(t)$ as:
\begin{equation}
\begin{split}
\theta_{j\rightarrow i}(t+\Delta t)=&(1-\theta_{j\rightarrow i}(t))
(1-\phi_{j\rightarrow i}(t))+\\
&(1-r_j(t)\Delta
t)\theta_{j\rightarrow i}(t) \;,
\label{cavityRate}
\end{split}
\end{equation}
where $\phi_{j\rightarrow i}(t)$ is the probability that node $j$ is not infected by any of its
neighbors excluding node $i$, which can be expressed as:
\begin{equation}
\phi_{j\rightarrow i}(t)= \prod_{\ell\in \mathcal{N}_{j}\setminus
  i}[1-\Delta t \, \widetilde{\lambda}_j(t)\, \theta_{\ell\rightarrow j}(t)].
\label{seccavity}
\end{equation}
The product in Eq.~(\ref{seccavity}) is over the set
$\mathcal{N}_{j}\setminus i$ of the neighbors of $j$ excluding
$i$. Further, by setting $\Delta t=1$ and considering situation in stationary state,
Eqs.~(\ref{dynamic}) and (\ref{cavityRate}) become
\begin{equation}
  \rho_i=(1-\rho_i)(1-\Omega_i)+(1-r_i)\rho_i,
  \label{steadyRho}
\end{equation}
and
\begin{equation}
  \theta_{j\rightarrow i}=(1-\theta_{j\rightarrow
    i})(1-\phi_{j\rightarrow i})+(1-r_j)\theta_{j\rightarrow i}.
  \label{steadyTheta}
\end{equation}
Through numerical iteration, we can compute the infection probability of any node at any
time $\rho_i(t)$, and prevalence $\rho$ in stationary state for different values
of $\alpha$ and $\lambda$. However, the equations can only be solved numerically,
except for the trivial solutions of $\rho_i=0$ and $\theta_{j\rightarrow i}=0$,
for all $i = 1,\cdots,N$, which leads to an overall $\rho=0$ phase of
an all-healthy population.

Due to nonlinearities in Eqs.~(\ref{resource})--(\ref{seccavity}), they
do not have a closed analytic form, and this disallows obtaining the
epidemic threshold $\lambda_c$ for fixed values of $\alpha$,
such that $\rho>0$ if $\lambda>\lambda_c$ and
$\rho=0$ when $\lambda<\lambda_c$. The calculation of $\lambda_c$
can be performed by considering that when $\lambda\rightarrow\lambda_c$,
$\rho_i\rightarrow0$ and $\theta_{j\rightarrow i}\rightarrow0$, and the
number of infected neighbors of any healthy node is
approximately zero in the thermodynamic limit.
Then prior to reaching $\lambda_c$, the expression
$(1-\theta_{j\rightarrow i})\rightarrow1$ is valid.
We can get a physical picture that the isolated infected
nodes are well-separated and surrounded by healthy nodes,
and any infected node $i$ will receive all the
resource from each of its neighbors.
By adding these assumptions to Eq.~(\ref{resource}),
resource $\omega_i$ becomes
$\omega_i= k_iq_0(1-\alpha)$. By linearizing Eq.~(\ref{reci})
and neglecting second-order terms for small $\mu$ , we obtain
\begin{equation}
\begin{aligned}
 r_i(t) &\approx \epsilon\mu\omega_i(t) \\
 &=\epsilon\mu k_iq_0(1-\alpha)
\end{aligned}
\label{linemu}
\end{equation}
Eq.~(\ref{linemu}) suggests that the recovery rate is
proportional to node degree and inversely proportional to
$\alpha$ when $\lambda\rightarrow\lambda_c$.
For the sake of clarity, the basic recovery rate is set at $\mu=0.01$
in this paper. Further the Eqs.~(\ref{q}) and (\ref{seccavity}) can also
be linearized using  $\theta_{j\rightarrow i} \approx 0$ as
\begin{equation}\label{cc}
  q_i\approx1-\widetilde{\lambda}_i\sum_{j=1}^{N}{a_{ji}\theta_{j\rightarrow
      i}}
\end{equation}
and
\begin{equation}
  \phi_{j\rightarrow i}\approx1- \widetilde{\lambda}_j\sum_{l\rightarrow h\in V_E}{\mathbf{M}_{j\rightarrow
      i,l\rightarrow h}\theta_{l\rightarrow h}},
\label{rseccavity}
\end{equation}
where $V_E$ is the set of directed edges and $\mathbf{M}$ is $|V_E|\times|V_E|$ non-backtracking matrix
\cite{krzakala2013spectral} of the network with the elements
labelled by the edges
\begin{equation}\label{nonbacking}
\mathbf{M}_{j\rightarrow i,l\rightarrow h}=\delta_{jh}(1-\delta_{il}),
\end{equation}
with $\delta_{il}$ being the Dirac delta function. Substituting
Eq.~(\ref{rseccavity}) into Eq.~(\ref{steadyTheta}) and ignoring higher order terms of $\theta_{j\rightarrow{i}}$ gives
\begin{equation}
 \sum{(-\delta_{lj}\delta_{ih}r_j+\widetilde{\lambda}_j\mathbf{M}_{j\rightarrow
    i,l\rightarrow h})}\theta_{l\rightarrow h}
    = 0
 \label{rcavityRate}
\end{equation}
Finally, consider that $m_j=1$ when $\lambda\rightarrow\lambda_c$,
Eq.~(\ref{lami}) becomes
\begin{equation}\label{lamj}
\widehat{\lambda}\equiv\widetilde{\lambda}_j=[(1-c)q_0(1-\alpha)+c]\lambda.
\end{equation}
To estimate the epidemic threshold, we calculate the average recovery rate as
\begin{equation}\label{aveRec}
  \langle{r}\rangle=\epsilon\mu\langle{k}\rangle q_0(1-\alpha)
\end{equation}
By inserting Eqs.~(\ref{lamj}) and (\ref{aveRec}) into Eq.~(\ref{rcavityRate}),
we get
\begin{equation}
 \sum{(-\delta_{lj}\delta_{ih}\langle{r}\rangle+\widehat{\lambda}\mathbf{M}_{j\rightarrow
    i,l\rightarrow h})}\theta_{l\rightarrow h}= 0
 \label{Finalrate}
\end{equation}
The system of equations in Eq.~(\ref{Finalrate}) has a non-trivial solution
if and only if $\langle{r}\rangle/\widehat{\lambda}$ is an eigenvalue of the matrix
$M$ \cite{gomez2010discrete}.  The lowest value $\widehat{\lambda}_c$
is then given by
\begin{center}
\begin{equation}
  \widehat{\lambda}_c=\frac{\langle{r}\rangle}{\Lambda_{max}} \, ,
\label{threshold}
\end{equation}
\end{center}
where $\Lambda_{max}$ is the largest eigenvalue of $\mathbf{M}$
\cite{shrestha2015message,pastor2015epidemic,wang2016unification}.

\section{Numerical verification and simulation results} \label{sec:simulation}
In this section, we will study the impact of systematically
human behavior on the dynamics of epidemic spreading and the
effects of network structure on the coupled dynamics of
resource allocation and disease spreadings respectively
through numerical verification and Monte Carlo simulations.
In the simulation, the synchronous updating
method~\cite{schonfisch1999synchronous,wang2015chaos} is applied to the disease infection and
resource allocation processes. Within each time increment $\Delta{t}$,
where $\Delta{t}=1$ in this paper, infection propagates from any I-state
node $j$ to S-state node $i$ with probability $\widetilde{\lambda}_i\Delta{t}$,
and any I-state node $j$ recovers to S-state with a probability $r_j\Delta{t}$.
With the spreading of disease, the resource allocation process co-occurs.
The dynamics terminate once it enters a steady state in which the number of infected nodes only fluctuates within a small range.
Note that, we fix the factor $c$ at a constant value $c=0.05$
throughout the paper, such that if any healthy individual $j$
chooses to reserve its resource, the probability that it is
infected in one contact with an infected neighbor
reduces to $\widetilde{\lambda}_j=0.05\lambda$.

\subsection{Effects of behavior response on the dynamics of epidemic spreading}
In this section, we investigate the effects of behavioral response of individuals on the spreading
dynamics. We consider that the coupled processes of resource allocation and
disease spreading takes place on a scale-free network, as many real-world networks
have skewed degree distributions
~\cite{girvan2002community,holme2002Growing,small2015growing,liu2003propagation}.
To build the network, we adopt the uncorrelated configuration
model(UCM) ~\cite{molloy1995critical,catanzaro2005generation} according to
a given degree distribution $P(k)\sim{k^{-\gamma}}$ with maximum degree
$k_{max}=\sqrt{N}$~\cite{boguna2004cut} and minimum degree $k_{min}=3$,
which assures no degree correlation of the
network when $N$ is sufficient large. To avoid the influence of network structure
on the result, the degree exponent is set at $\gamma=2.4$, the network size is set
at $N=10000$, and average degree is set at $\langle{k}\rangle=8$ in the simulations.
In addition, we leverage the susceptibility measure $\chi$ to
determine the epidemic threshold through simulations~\cite{ferreira2012epidemic},
which is expressed as
\begin{equation}
\chi = N\frac{\langle\rho^2\rangle-\langle\rho\rangle^2}{\langle\rho\rangle},
\end{equation}
where $\langle\ldots\rangle$ is the ensemble averaging.
The epidemic threshold can then be determined when the value of $\chi$
exhibits diverging peaks at the certain infection rate
\cite{ferreira2012epidemic,chen2016crossover}.

\begin{figure}
  \centering
  \includegraphics[width=1\linewidth]{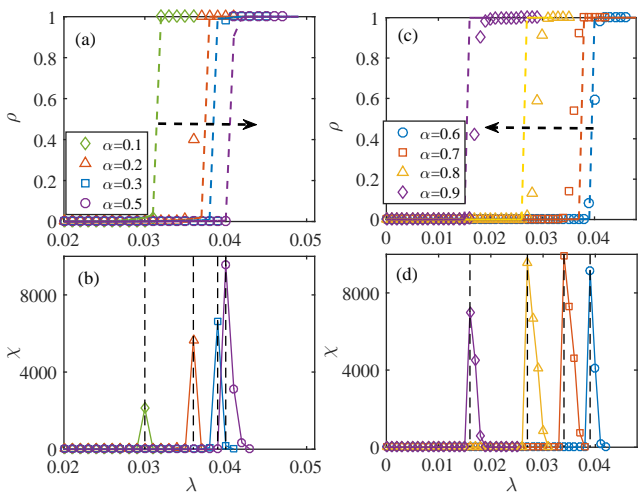}\\
  \caption{Effects of behavior response on the dynamics of disease spreading on a scale-free network. (a) and (c): The prevalence $\rho$ in stationary
   state as a function of basic infection rate $\lambda$ for
  varieties of reaction strength $\alpha$. Symbols represent the results obtained from Monte Carlo simulations, and lines represent the results of the GDMP method.
  (b) and (d): The corresponding susceptibility measure $\chi$
  as a function of $\lambda$. Data are obtained by
  averaging over 500 independent simulations.}
  \label{infsize_sf}
\end{figure}
We first investigate the effects of behavior response on the spreading dynamics using Monte Carlo simulations. Initially, a fraction of $\rho(0)=0.1$ nodes
are selected randomly as seeds, the remain nodes are in susceptible state. To present different reaction strength of
individuals when they are aware of a certain disease from local
information, we select eight typical values of $\alpha$ from
$\alpha=0.1$ to $\alpha=0.9$ in the simulation. In Figs.~\ref{infsize_sf}
(a) and (c), we plot the prevalence $\rho$ in the stationary state as a
function of basic infection rate $\lambda$ for different $\alpha$.
Symbols in Figs.~\ref{infsize_sf} (a) and (c) represent the
results obtained by Monte Carlo simulations and lines are the
theoretical results obtained from numeric iterations, respectively.
From the curves Figs.~\ref{infsize_sf}(a) and (c), we observe
that the system converges to two possible stationary states: either
the whole population is healthy, or it becomes completely infected
for any $\alpha$, which tells us that when there is a shortage in
resource, the disease will break out abruptly.

Besides we can observe from Figs.~\ref{infsize_sf}(a) and (b)
that with the increase of $\alpha$ from $\alpha=0.1$ to $\alpha=0.5$,
the epidemic threshold increases gradually, see the peaks of $\chi$
for the corresponding $\alpha$. It reveals that
the stronger the individual's sense of self-protection,
the more delayed the outbreak of the disease within this parameter interval
(see the right arrow).
On the contrary, we observe from Figs.~\ref{infsize_sf}(c) and (d) that when
$\alpha$ increases from $\alpha=0.6$ to $\alpha=0.9$, the
threshold decreases gradually, which reveals that
the disease breaks out more easily with a stronger sense of self-protection
within this parameter interval (see the left arrow).
The phenomenon suggests that too cautious or too selfless for the people
during the outbreak of an epidemic are both not suitable for disease control, and
there is an optimal value of the reaction strength, at which an epidemic
outbreak will be postponed to the greatest extent.

\begin{figure}
  \centering
  \includegraphics[width=1\linewidth]{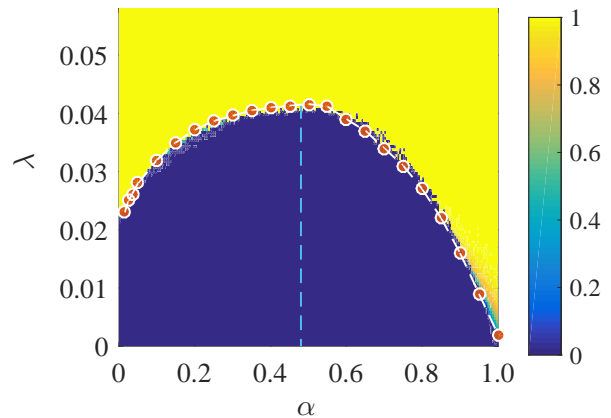}\\
  \caption{The phase diagram in parameter plane $(\alpha-\lambda)$ on scale-free network. Colors encode the value of $\rho$ obtained from Monte Carlo simulations. Red circles connected by dotted lines represent  theoretical predictions of epidemic threshold $\lambda_c$. The blue dotted line indicates the location of optimal value $\alpha_{opt}$. Data are obtained by averaging 50
  Monte Carlo simulations for each point in the grid $200\times200$}
  \label{SFinf3d}
\end{figure}
We further study systematically the effects of behavior response and basic infection
rate on the spreading dynamics. In Fig.~\ref{SFinf3d}, we exhibit the full phase diagram
$(\alpha-\lambda)$ of the coupled dynamics of resource allocation and disease spreading.
Colors in Fig.~\ref{SFinf3d} (a) encode the fraction of infected nodes in the stationary
state $\rho$. The epidemic threshold $\lambda_c$, marked by red circles,
rises monotonically until it reaches the maximum at $\alpha_{opt}$ (indicated by the
blue dotted line), and then falls gradually with the increase of $\alpha$.
Besides, we observe that there are only two possible
stationary state: the whole healthy (marked by blue color)
and the whole infected of the population (marked by yellow color),

\begin{figure}
  \centering
  \includegraphics[width=1\linewidth]{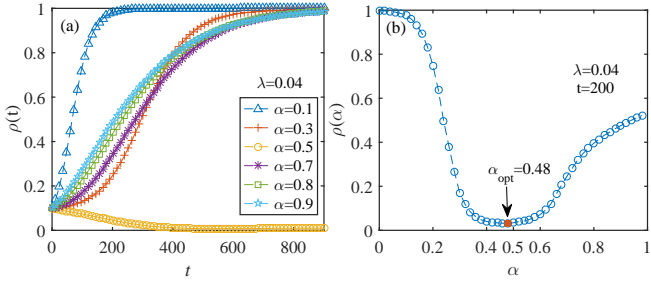}\\
  \caption{Effects of behavior response on evolution of the fraction of
  infected nodes $\rho(t)$. (a) The time evolution of $\rho(t)$
  for varieties of $\alpha$ using Monte Carlo simulations for a fixed
  value of $\lambda=0.04$. (b) Plot of the fraction of infected nodes
  versus the change in $\alpha$ at a fixed time $t=200$ and infection rate
  $\lambda=0.04$. The results of the simulations are obtained by averaging
  over 300 realizations. }\label{timeRho}
\end{figure}
Fig.~\ref{timeRho} (a) plots the time evolution of $\rho(t)$ for
six typical values of $\alpha$ when the basic infection
rate is fixed at $\lambda=0.04$. We find that when the value of
$\alpha$ is small, the system to converge to a stationary state rapidly,
such as $\rho(\infty)=1.0$ for $\alpha=0.1$. With the increase of
$\alpha$, it takes a longer time for the system to reach a stationary
state. Further, to exhibit the effects of $\alpha$ on the dynamics more intuitively,
we plot the fraction of infected nodes at a fixed time $t=200$ as
a function of $\alpha$ in Fig.~\ref{timeRho} (b),
which is denoted as $\rho(\alpha)$ for the sake of clarity.
We observe that the value of $\rho(\alpha)$ decreases continuously with
$\alpha$ until reaching the minimum value at $\alpha_{opt}\approx0.48$
(marked by red circle in Fig.~\ref{timeRho}(b)), and
then increases gradually with $\alpha$.

\begin{figure}
  \centering
  \includegraphics[width=1\linewidth]{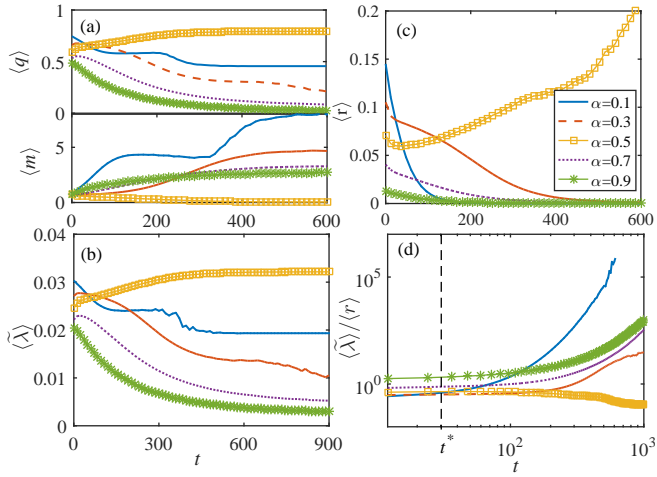}\\
  \caption{ Plots of the critical parameters versus $t$ for typical values
  of $\alpha$. (a) Top pane: Time evolution of average donation rate $\langle{q}\rangle$.
  Bottom pane: The evolution of average number of infected neighbors
of all nodes $\langle{m}\rangle$.
(b) Time evolution of average infection rate $\langle{\widetilde{\lambda}}\rangle$.
  (c) The complete evolution of average recovery rate $\langle{r}\rangle$.
  (d) Log-log plots of average effective infection rate $\langle{\widetilde{\lambda}}\rangle/\langle{r}\rangle$.
basic infection rate is fixed at $\lambda=0.04$. The results of the
simulations are obtained by averaging over 300 realizations }\label{timePara}
\end{figure}
Next we qualitatively explain the optimal phenomena by studying the time evolution
of the critical quantities.

We begin by studying the case when $\alpha$ is small, for example, $\alpha=0.1$.
We observe in Fig.~\ref{timePara} that in the initial stage,
the donation probability for $\alpha=0.1$ is the highest [see the blue line
in the top panel of Fig.~\ref{timePara} (a)], since
a smaller value of $\alpha$ means a higher willingness of
healthy individuals to allocate resources.
Although the resource of healthy individuals can
improve the recovery probability of infected neighbors to
a certain extent, it also makes themselves more likely to be infected.
We can observe in Figs.~\ref{timePara} (b) and (c) that, the average
recovery rate $\langle{r}\rangle$ and infection rate
$\langle{\widetilde{\lambda}}\rangle$ is highest for $\alpha=0.1$, meanwhile, there is
a lowest value of effective infection rate $\langle{\widetilde{\lambda}}\rangle/\langle{r}\rangle$, as shown in Fig.~\ref{timePara} (d).
However, as the high probability of being infected for the healthy nodes,
the number of infected individuals increases at a high rate [see the blue line
in the bottom pane of Fig.~\ref{timePara} (a)]. When people are aware
of the increment of the infected neighbors, they will reduce
their donation willingness, which leads to a reduction in
infection rate $\langle{\widetilde{\lambda}}\rangle$, as shown
in Figs.~\ref{timePara} (a) and (b). Consequently, with less
resource received from healthy neighbors, the recovery rate of
infected nodes reduces accordingly, see Fig.~\ref{timePara} (c),
which leads to an increase of the effective infection
rate $\langle{\widetilde{\lambda}}\rangle/\langle{r}\rangle$~\cite{pastor2015epidemic}.
The increase in $\langle{\widetilde{\lambda}}\rangle/\langle{r}\rangle$
has led to a further increase in the number of infected nodes.
Then people become more aware of the threat of disease,
and thus reduce the probability of resource donation further,
which leads to a further decrease in infection rate
$\langle{\widetilde{\lambda}}\rangle$ and recovery rate
$\langle{r}\rangle$ and finally, the increase of the
effective infection rate $\langle{\widetilde{\lambda}}\rangle/\langle{r}\rangle$.

Specifically, we observe from Fig.~\ref{timePara} (d) that,
when it surpasses a critical time $t^*$, indicated
by the dotted line in the figure, the value of $\langle{\widetilde{\lambda}}\rangle/\langle{r}\rangle$
proliferates, which suggests that in this stage the
infection of healthy individuals is much faster than
the recovery of infected individuals.
With more newly infected nodes, the donation probability $\langle{q}\rangle$
and infection rate $\langle{\widetilde{\lambda}}\rangle$ decreases further,
which results in less resources donated to support the recovery of infected nodes.
Thus the recovery rate of infected nodes $\langle{r}\rangle$ drops abruptly,
which in turn promotes the increases of effective infection rate
$\langle{\widetilde{\lambda}}\rangle/\langle{r}\rangle$ further, and
then more and number of infected nodes appear.
Consequently, the cascading failure of the entire system occurs.

Based on the above analysis for a small value of $\alpha$, i.e., $\alpha=0.1$,
we can reasonably explain why people are more willing to contribute a resource
while the disease is more likely to break out.

Secondly, we study the case when $\alpha$ is significant, for example, $\alpha=0.9$.
As a larger value of $\alpha$ means more sensitive of the individuals
to the disease and a lower willingness to allocate resources. Thus
we observe from Fig.~\ref{timePara} (a) that, initially, there
is a smallest value of $\langle{q}\rangle$ [see the green
stars in top pane of Fig.~\ref{timePara} (a)], and infection
rate $\langle{\widetilde{\lambda}}\rangle$. Consequently,
the infected nodes will receive the lowest value of the resource
to recover, which leading to the smallest value of recovery rate
$\langle{r}\rangle$, as shown the green stars in of Fig.~\ref{timePara} (c).
Then the recovery of infected nodes is delayed leading to a high
effective infection rate. We can observe in Fig.~\ref{timePara} (d) that,
when $\alpha=0.9$, there is a highest value of
$\langle{\widetilde{\lambda}}\rangle/\langle{r}\rangle$.
The high effective infection rate will lead to a rapid increase in
the number of infected nodes. We can observe in the bottom pane of
Fig.~\ref{timePara} (a) that, in the early stage, there is a second
largest value of $\langle{m}\rangle$ for $\alpha=0.9$, as denoted by
the green stars. The large value of $\langle{m}\rangle$ will
further, reduce the willingness of resource donation for the
healthy individuals, thus we can observe a continuous decline in
$\langle{q}\rangle$ and $\langle{\widetilde{\lambda}}\rangle$.
What's worse, the recovery rate of infected nodes
keeps declining with less and less resource
[see the curve in Fig.~\ref{timePara} (c)], which leading to
a rapid growth of $\langle{\widetilde{\lambda}}\rangle/\langle{r}\rangle$
[see the curve in Fig.~\ref{timePara} (d)].

Thus we can explain the reason why a higher sense of
self-protection of the population can not suppress the
disease effectively.

At last, we observe in Fig.~\ref{timePara} that, when the
value of $\alpha$ is around the optimal value $\alpha_{opt}$,
there is a relatively lower value of $\langle{q}\rangle$
comparing to the case of $\alpha=0.1$ in the initial stage,
which results in a lower value of $\langle{\widetilde{\lambda}}\rangle$
[see the yellow squares in Figs~\ref{timePara} (a) and (b)].
The lower willingness of resource donation induces to a relatively
smaller value of recovery rate $\langle{r}\rangle$, as shown
in Fig.~\ref{timePara} (c). However, we can observe from Fig.~\ref{timePara} (d)
that the effective infection rate
$\langle{\widetilde{\lambda}}\rangle/\langle{r}\rangle$
keeps the lowest value in the early stage, which suggests that
the disease will propagate slowly in the population, and
the number of infected nodes will increase slowly, which
is verified by the curve in the bottom pane of Fig.~\ref{timePara} (a).
Further, the small value of $\langle{m}\rangle$ will promotes the
increase of $\langle{q}\rangle$ [see the curve in the top pane of Fig.~\ref{timePara}
(a)], which resulting in the increase of recovery rate $\langle{r}\rangle$.
And finally, the effective infection rate $\langle{\widetilde{\lambda}}\rangle/\langle{r}\rangle$ decreases further,
as shown in Fig.~\ref{timePara} (d). Thus the disease can be
suppressed to the greatest extend.
\begin{figure}
  \centering
  \includegraphics[width=1\linewidth]{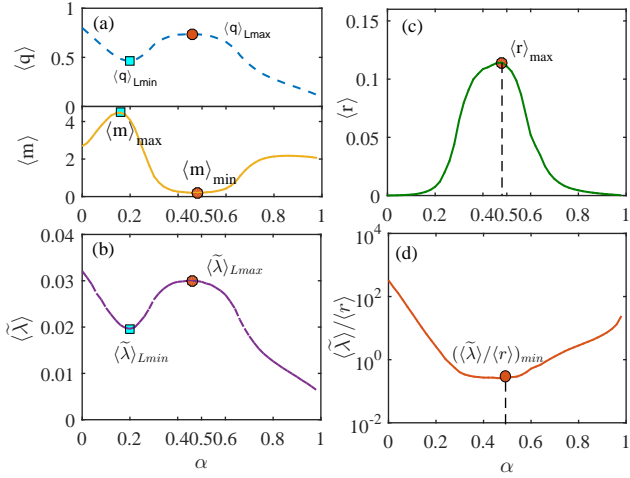}\\
  \caption{Plots of the critical parameters versus $\alpha$ at fixed time $t=200$
  , and basic infection rate $\lambda=0.04$.
  (a) Top pane: the average donation rate $\langle{q}\rangle$ as a function of $\alpha$.
  Bottom pane: The average number of infected neighbors of all nodes $\langle{m}\rangle$
  as a function of $\alpha$.
(b) Average infection rate $\langle{\widetilde{\lambda}}\rangle$ as a function of $\alpha$.
  (c) The average recovery rate $\langle{r}\rangle$ as a function of $\alpha$.
  (d) Plots of average effective infection rate $\langle{\widetilde{\lambda}}\rangle/\langle{r}\rangle$.
as a function of $\alpha$. The results of the
simulations are obtained by averaging over 300 realizations}\label{timeAlphaPara}
\end{figure}
Through the three steps, we explain the optimal phenomena in the coupled
dynamics of resource allocation and disease spreading.

Finally, we further verify our explanation by studying the critical quantities
as the function of parameter $\alpha$ at a fixed time $t$
and basic infection rate $\lambda$. Figs.~\ref{timeAlphaPara} (a)
to (d) plot the value of $\langle{q}\rangle$, $\langle{m}\rangle$,
$\langle{\widetilde{\lambda}}\rangle$, $\langle{r}\rangle$ and
$\langle{\widetilde{\lambda}}\rangle/\langle{r}\rangle$
as a function of $\alpha$ when $t=200$, and $\lambda=0.04$.
For the sake of clarity,
we denote the local minimum and maximum value as $X_{Lmin}$,
$X_{max}$, and the global minimum and maximum value as
$X_{min}$ and $X_{max}$ respectively, where $X\in[\langle{q}\rangle,
\langle{m}\rangle,\langle{\widetilde{\lambda}}\rangle,
\langle{\widetilde{\lambda}}\rangle/\langle{r}\rangle]$.
We observe that, although when $\alpha$
is around $\alpha_{opt}$, there is a local maximum of
$\langle{q}\rangle_{Lmax}$ and $\langle{\widetilde{\lambda}}\rangle_{Lmax}$.
The recovery rate reaches maximum $\langle{q}\rangle_{max}$, and the
effective infection rate reaches lowest $(\langle{\widetilde{\lambda}}\rangle/\langle{r}\rangle)_{min}$,
which indicates that the disease can be optimally suppressed at
this point.

\subsection{Effects of network structure on coupled dynamics}
In this section, we investigate the effects of network structure on
the coupled dynamics of resource allocation and disease spreading.
To avoid the impact of reaction strength on the result, the parameter
$\alpha$ is fixed at $\alpha=0.5$. In addition, we adopt the UCM model
to generate scale-free networks with different degree distributions
$P(k)\sim{k}^{-\gamma}$. As the degree heterogeneity decreases with
the increase of the power exponent $\gamma$ \cite{boccaletti2006complex,newman2010networks},
thus it approaches to random regular networks $RRNs$ when $\gamma\rightarrow{\infty}$ \cite{chen2018}.

\begin{figure}
  \centering
  \includegraphics[width=0.9\linewidth]{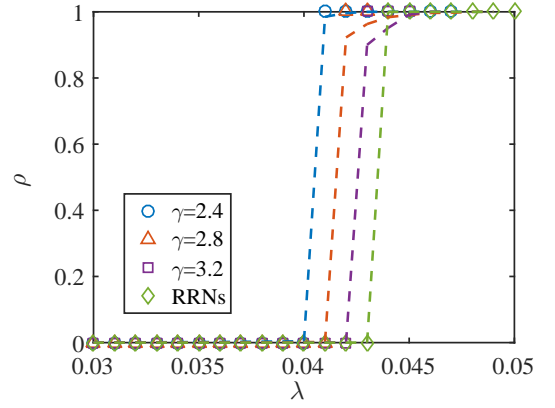}\\
  \caption{The prevalence $\rho$ in the stationary as a function of
  $\lambda$ on scale-free networks with degree exponent $\gamma=2.4$ (blue circles),
  $\gamma=2.8$ and $\gamma=3.2$ (purple squares). And the result on
  random regular network (RRNs) marked by red rhombus. Symbols represent
  the results obtained from  Monte Carlo simulations, and lines represent results of GDMP method. The parameter $\alpha$ is fixed at $\alpha=0.5$}\label{net}
\end{figure}

Fig.~\ref{net} plots the prevalence $\rho$ in the stationary state as a function of
the basic infection rate $\gamma$ for networks with four typical values of $\gamma$: $\gamma=2.4$ (blue circles), $\gamma=2.8$ (upper triangles), $\gamma=3.2$ (purple squares) and $\gamma\rightarrow\infty$ (red rhombus). We observe that there are
only two stationary states of the system: all healthy or completely infected
for all networks, which implies that the network structure does not alter the
first-order transition of $\rho$. Besides, we find that, with an increase
of $\gamma$, the outbreak of disease is delayed gradually. It suggests that
the degree heterogeneity enhance the disease spreading, which is
consistent with the existing research conclusions \cite{pastor2001epidemicprl}.

\begin{figure}
  \centering
  \includegraphics[width=1\linewidth]{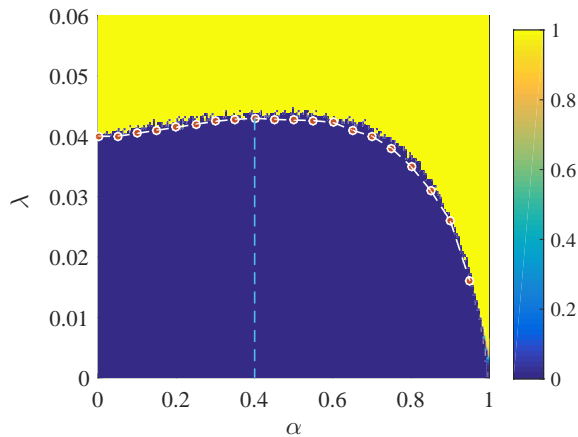}\\
  \caption{The phase diagram in parameter plane $(\alpha-\lambda)$ on RRNs. Colors encode the value of $\rho$ obtained from Monte Carlo simulations. Red circles connected by dotted lines represent theoretical predictions of epidemic threshold $\lambda_c$. The blue dotted line indicates the location of optimal value $\alpha_{opt}$. Data are obtained by averaging 50
  Monte Carlo simulations for each point in the grid $200\times200$ }\label{RRinf3d}
\end{figure}
\vspace{10pt}

In the end, we study the effects of behavior response on the spreading
dynamics on the spreading dynamics systematically. Fig.~\ref{RRinf3d} is the phase diagram in parameter plane $(\alpha-\lambda)$ on RRNs. Colors encode the prevalence in the
stationary state $\rho$. We find that there is also an optimal value $\alpha_{opt}$,
at which the epidemic threshold reaches the maximum, indicated by the blue dotted line
if Fig.~\ref{RRinf3d}. The results suggest that the network structure does not
alert the optimal phenomenon in behavior response.
\section{Discussion} \label{sec:dis}
In this paper, we focus on the problem of how can we protect ourselves from
being infected while helping others by donating resources during an outbreak of an epidemic.
To answer this question, we propose a novel resource allocation model in controlling
the epidemic spreading. We consider that healthy individual can contribute their
resources to help those in need, and some will consider self-protection first when they
perceive the threat of disease. The others will contribute to resource as much as possible.
To quantify the behavior response of individuals, a tune parameter $\alpha$ is introduced.
Besides, to study the coupled dynamics of resource allocation and disease spreading,
a resource-based SIS model is proposed. First of all, we theoretically analyze the model
using a generated dynamic message-passing method, and then we carry out extensive
Monte Carlo simulations on both scale-free and random regular networks.
Through theoretical analysis and simulations, we find that the coupled dynamics converges
to two stationary states: the whole infected or all healthy, which indicates that a
shortage of resource can induce an abrupt outbreak of the epidemic. More importantly,
we find that too cautious or too selfless for the people during the outbreak of
an epidemic are both not suitable for disease control. There is an optimal (balance) point
where the epidemic spreading can be controlled to the greatest extent.
It also tells us that we can donate resource appropriately to support the people in need,
but at the same time, we should keep some resources for self-protection.
Further, we find out the optimal point on both in certain conditions.
At last, we investigate the effects of network structure on the coupled dynamics.
We find that the degree heterogeneity promotes the outbreak of disease,
and the network structure does not alter the optimal phenomenon in behavior
response.

The discovery of the optimal (balance) point
is of practical significance for controlling the outbreak of
infectious diseases, especially in the context of the outbreak of 2019-ncov,
in Wuhan, China. It will guide us
to make the most reasonable choice between resource contribution and self-protection
when perceiving the threat of disease.

\section{Acknowledgements}
This work was supported by the Fundamental Research Funds for the Central Universities
(Nos. JBK190972, JBK171113, JBK170505),
National Natural Science Foundation of China (Nos.61903266,71671141,71873108),
the Financial Intelligence \& Financial Engineering Key Lab of Sichuan Province,
and China Postdoc-toral Science Foundation (No. 2018M631073),
China Postdoctoral Science Special Foundation (No. 2019T120829).
\section*{References}

\bibliography{mybibfile}
\end{document}